\documentclass{article}

\usepackage{amsmath,amsfonts,amssymb,amsthm,cleveref,color}
\usepackage{enumerate}
\usepackage{ifthen}

\usepackage[]{color}

\newtheorem{theorem}{Theorem}
\newtheorem{definition}[theorem]{Definition}

\newtheorem{examples}[theorem]{Examples}

\newtheorem{proposition}[theorem]{Proposition}

\begin{document}
\title{Courcelle's Theorem without Logic}
\author{Yuval Filmus and Johann A. Makowsky \\ \small Faculty of Computer Science \\ \small Technion-Israel Institute of Technology, Haifa, Israel}

\maketitle
\begin{abstract}
Courcelle's Theorem states that
on graphs $G$ of tree-width at most $k$ with a given tree-decomposition of size $t(G)$, 
graph properties $\mathcal{P}$ definable in Monadic Second Order Logic
can be checked in linear time in the size of $t(G)$.
Inspired by L. Lov\'asz' work using connection matrices instead of logic,
we give a generalized version of Courcelle's theorem which replaces the definability hypothesis
by a purely combinatorial hypothesis using a generalization of connection matrices.

\end{abstract}

\newif\ifskip
\skiptrue
\newcommand{\cH}{\mathcal{H}}
\newcommand{\cC}{\mathcal{C}}
\newcommand{\cD}{\mathcal{D}}
\newcommand{\cE}{\mathcal{E}}
\newcommand{\cP}{\mathcal{P}}
\newcommand{\cT}{\mathcal{T}}
\newcommand{\fA}{{\mathfrak A}}
\newcommand{\fB}{{\mathfrak B}}
\newcommand{\fC}{{\mathfrak C}}
\newcommand{\N}{{\mathbb N}}
\newcommand{\Z}{{\mathbb Z}}
\newcommand{\MSOL}{\mathbf{MSOL}}
\newcommand{\SOL}{\mathbf{SOL}}
\newcommand{\CMSOL}{\mathbf{CMSOL}}
\newcommand{\FOL}{\mathbf{FOL}}
\newcommand{\FPT}{\mathbf{FPT}}
\newcommand{\NP}{\mathbf{NP}}
\newcommand{\Str}{\mathrm{Str}}
\newcommand{\MW}{\mathrm{MW}}
\newcommand{\CW}{\mathrm{CW}}
\newcommand{\TW}{\mathrm{TW}}
\newcommand{\TWW}{\mathrm{TWW}}
\newcommand{\bF}{\mathbf{F}}
\newcommand{\bA}{\mathbf{A}}

\section{Introduction and outline}
\label{se:intro}
We assume the reader is familiar the notion of tree-width and related width notions
in graph theory, with Courcelle's Theorem, 
and with the notion of fixed prameter complexity.
Good references are
(\cite{hlinveny2008width,courcelle2012graph,downey2012parameterized,downey2013fundamentals}). 

\subsection{Tree decompositions}

A tree decomposition of a graph $G$  gives a natural notion of width, the {\em tree-width, denoted by $tw(G)$}.

Courcelle's Theorem states:

\begin{theorem}[Courcelle]
\label{th:courcelle}
Let $\cC$ be a class of graphs of tree-width at most $k$ and $\phi$ be a formula of $\MSOL$.
Then checking whether a graph $G \in \cC$ satisfies $\phi$ is in $\FPT$.
\end{theorem}

A similar theorem holds for graph polynomials:

\begin{theorem}[Courcelle, Makowsky, Rotics]
\label{th:CMR}
Let $\cC$ be a class of graphs of tree-width at most $k$ and $F(G; \bar{x})$ be a $\MSOL$-definable graph polynomial.
Then evaluating $F(G; \bar{x})$ for graphs $G \in \cC$ is in $\FPT$.
\end{theorem}

A simple univariate $\MSOL$-definable graph polynomial is of the form
\begin{gather}
F(G; x) = \sum_{A \subset V(G) : G[A] \models \phi } x^{|A|}
\end{gather}
where $G[A]$ is the induced subgraph of $G$ with vertex set $A$ and $G[A] \models \phi$.

One way of proving Theorems \ref{th:courcelle} and \ref{th:CMR}   makes extensive use of two ingredients:
\begin{enumerate}[(i)]
\item
The class of of graphs of tree-width at most $k$ has an inductive definition using $k$-connections of graphs with $k$ ports, and
\item
the Feferman-Vaught Theorem for $\MSOL$.
\end{enumerate}

\subsection{The Feferman-Vaught Theorem}
The Feferman-Vaught Theorem is a global theorem which applies to all formulas of $\FOL$. The version for $\MSOL$ 
was
proved by H. L\"auchli and, independently by S. Shelah 
\cite{lauchli1968decision,ar:shelah75} and 
for $\CMSOL$ by B. Courcelle \cite{courcelle1990monadic}. For a history, survey, and many applications see \cite{makowsky2004algorithmic}.

In \cite[Theorem 4.21]{makowsky2004algorithmic} an abstract version of Theorem \ref{th:courcelle} is proved based on two ingredients alone.
\begin{enumerate}[(i)]
\item
A class of of graphs $\cC$ inductively defined using  finitely many $\MSOL$-smooth ($\CMSOL$-smooth) operations. 
A {\em parse-tree $t(G)$} for $G$ encodes the way $G \in \cC$ was inductively defined. The size of the parse-tree  $t(G)$ is its number of nodes.
\item
The fact that there are for every $k, q \in \N^+$, up to logical equivalence, only finitely many $\MSOL$ ($\CMSOL$) formulas  in $k$ free variables
and quantifier rank at most $q$.
\end{enumerate}
An $n$-ary operation $\Box$ on $\tau$-structures is $\MSOL$-smooth ($\CMSOL$-smooth) if for all $\fA_1, \ldots \fA_n$, $\fB_1, \ldots \fB_n$
such that for all $i \in [n]$ the structures $\fA_i$ and $\fB_i$ satisfy  pairwise the same $\MSOL$ sentences of quantifier rank $q$,
then $\Box(\fA_1, \ldots \fA_n)$ and $\Box(\fB_1, \ldots \fB_n)$ also satisfy the same $\MSOL$ ($\CMSOL$) sentences of quantifier rank $q$.
The Feferman-Vaught Theorem is a consequence of smoothness. 

\begin{theorem}[Courcelle, Makowsky]
\label{th:CM-abstract}
Let $\cC$ as above and let $G \in \cC$ with a parse-tree $t(G)$ of size $m(G)$ and $\cP$ be a $\CMSOL$-definable graph property.
Then checking $G \in \cP$ is in linear time  in the size of the parse-stree $m(G)$.
\end{theorem}

This theorem seems like the possibly most general version of Courcelle's theorem which does use definability and smothness for $\CMSOL$.
It does cover the cases of clique-width and modular-width. 

This paper shows how we can avoid definability in $\CMSOL$ and prove a purely combinatorial theorem similar to Theorem \ref{th:CM-abstract}.

\subsection{Connection matrices and Hankel matrices}
If one is only interested in a particular graph property $\cP$ defined by some formula $\phi \in \MSOL$,
it was noticed by L. Lov\'asz that one can replace the graph property defined by $\phi \in \MSOL$ by a graph property $\cP$
such that 
its connection matrices have finite rank.
Connection matrices were introduced in unpublished preprints in 2003 and 2005
\cite{freedman2003graph,freedman2005graph}
and used in the published version \cite{lovasz2007connection,freedman2007reflection}.
They are special cases of Hankel matrices defined for binary operations on finite structures, in particular graphs, labeled graphs or colored graphs.
In \cite[Chapters 4-6]{lovasz2012large} Lov\'asz used them in order to prove the following:

\begin{theorem}\cite[Theorem 6.48]{lovasz2012large}
\label{th:lovasz}
Let $\cC$ be a class of graphs of tree-width at most $k$ and $\cP$ be a graph property
such that its $k$-connection matrix has finite rank.
Then checking whether a graph $G \in \cC$ with parse-tree $t(G)$  satisfies property $\cP$ is in $\FPT$ in the size of $t(G)$.
\end{theorem}

A $k$-connection of two graphs $G_1, G_2$ with each having $k$ distinguished elements (aka ports) $a_i^1, \ldots , a_k^i, i=1,2$ is 
the disjoint union of $G_1, G_2$ denoted by $G_1 \sqcup_k G_2$ where the  corresponding elements $a_j^1$ and $a_j^2$ are identified.
Connection  matrices are special cases of Hankel matrices restricted to $k$-graphs and $k$-connections. 

Let $\tau$ be a finite vocabulary and let $\Str(\tau)$ be the set of all $\tau$-structures with universe some $[n]$.
Let $\Box$ be a binary operation on $\Str(\tau)$ preserved under $\tau$-isomorphism and let $\cP$ be a  
property of $\tau$-structures.
Let $(\fA_i, i \in \N)$ be an enumeration of $\Str(\tau)$.
For $i=0$ we put the empty structure, i.e. the structure with the empty set as its universe.

We define an infinite $(0,1)$-matrix $\cH(\Box, \cP)$ as follows:
\begin{gather}
\cH(\Box, \cP)_({i,j}) =
\begin{cases}
1 & \fA_i \Box \fA_j \in \cP \\
0 & \text{otherwise}
\end{cases}
\end{gather}
We denote by $r(\Box, \cP)$ the rank of $\cH(\Box, \cP)$ over $GF(2)$.
We shall define and discuss connection matrices and Hankel matrices in more detail  in Section \ref{se:hankel}.
There we also show that that there are continuum many graph properties whose connection matrix has rank $2$. 
Hence, Theorem \ref{th:lovasz} has a much larger range of applicability, since there are only countably many graph properties
definable in $\MSOL$.

A simple graph polynomial without a definability assumption is of the form
\begin{gather}
F(G; x) = \sum_{A \subset V(G) : G[A] \in \cP} x^{|A|}.
\end{gather}
A general defintion is given in \cite{kotek2011counting,godlin2008evaluations}, but is not needed here.
There $\cP$ is assumed to be definable in $\CMSOL$. Here we do not assume definability.

The theorem corresponding to Theorem \ref {th:CMR} for graph polynomials can be formulated and proved as follows:
\begin{theorem}[Lov\'asz]
\label{th-lovasz-1}
Let $\cC$ be a class of graphs of tree-width at most $k$ and $\cP$ be a graph property
such that its connection matrices have finite rank.
Let $G \in \cC$ with parse-tree $t(G)$.
\begin{gather}
F(G; x) = \sum_{A \subset V(G) : G[A] \in \cP} x^{|A|}.
\end{gather}
Then evaluating $F(G; x)$ for graphs $G \in \cC$ is in $\FPT$ in the size of $t(G)$.
\end{theorem}
The proof is technical but straight forward and combines the methods from 
\cite{makowsky2004algorithmic} and \cite{lovasz2012large}.
However, one has to be careful in the choice of the inductive definition of graphs of tree-width at mosr $k$. 

In the proof of Theorem \ref{th:CM-abstract} it is important that the notion of $\MSOL$-smooth and $\CMSOL$-smooth
is well defined for graph operations of arbitrary arity. In contrast, Hankel matrices are defined only for binary operations.

\subsection{The combined circuit matrix and main result}
Let $\cC$ be a class of $\tau$-structures defined inductively using a fixed finite set of 
$\tau$-structures $\bA = \{\fA_i: i \in [a]\}$ and a finite set $\bF$ of
operations on $\tau$-structures $\bF = \{F_i: i \in [b]\}$ of arity $r_i: i \in [b]$ as follows:
\begin{enumerate}[(i)]
\item
$\bA \subseteq \cC$.
\item
If $F_j \in \bF$ and $\fB_i \in \cC$ for $i \in [r_j]$ 
then $F_j(\fB_1, \ldots \fB_{r_j}) \in \cC$.
\end{enumerate}
Let $t(\fA)$ be a parse-tree for $\fA \in \cC$.

Hankel matrices are only defined for binary operations of $\tau$-structures.
If an inductively defined classes of graphs is defined using more than one binary operation it is not obvious that assuming
that for each binary operation the corresponding Hankel matrix is of finite rank is enough.
We want to have an abstract theorem for inductively defined classes of graphs for general inductive definitions.

In order to remedy this we define a {\em combined matrix $\cH(\bA, \bF, \cP)$} based on all the operations needed to 
generate 
inductively
the graphs of $\cC$ with entries in $\{0,1\}$ and a class of $\tau$-structures $\cP$.
If $\bF$ contains only one binary operation $\Box$ and $\bA$ is empty, the resulting combined matrix 
$\cH(\bA, \bF, \cP)$ for $\cP$ is just the  Hankel matrix $\cH(\Box, \cP)$.
More generally we have

\begin{proposition}
\label{pr:hankel-2}
If $\Box \in \bF$ is a binary operation on $\tau$-structures,  $\cH(\Box, \cP)$ is a submatrix of $\cH(\bA, \bF, \cP)$. 
Furthermore, if $\cH(\bA, \bF, \cP)$ has rank $r$, then $\cH(\Box, \cP)$ has rank $\leq r$.
\end{proposition}
We shall prove this in Section \ref{se:circuits}.

Our main result can now be stated as follows:

\begin{theorem}
\label{th:main}
Let $\cC$ be a class of $\tau$-structures defined inductively using $\bA, \bF$.
Assume the combined matrix
$\cH(\bA, \bF, \cP)$ is of finite rank. 
For $\fA \in \cC$ with parse-tree $t(\fA)$, checking whether $\fA \in \cP$ can be done in linear time
in the size of $t_{\fA}$ where the constants depend only on the rank of $\cH(\bA, \bF, \cP)$. 
\end{theorem}

This can be viewed as a generalization of Lov\'asz' Theorem (Theorem \ref{th:lovasz}) because of Proposition \ref{pr:hankel-2}.


\subsection{Outline of the paper}

In Section \ref{se:hankel}
we discuss Hankel matrices in detail and (possibly) give a full proof of Theorem \ref{th:lovasz}.
In Section \ref{se:circuits} we define the combined circuit  matrix and prove its basic properties,
and prove our main result, Theorem \ref{th:main}, and in
Section \ref{se:applications} discuss some of its applications.
In Section \ref{se:conclu} we discuss the meaning of our result and state directions for further
research.

\section{Hankel matrices}
\label{se:hankel}

Let $\tau$ be a finite vocabulary and let $\Str(\tau)$ be the set of all $\tau$-structures with universe some $[n]$.
Let $\Box$ be a binary operation on $\Str(\tau)$ preserved under $\tau$-isomorphism and let $\cD$ be a $\tau$-property.
Let $(\fA_i, i \in \N)$ be an enumeration of $\Str(\tau)$.
For $i=0$ we put the empty structure, i.e. the structure with the empty set as its universe.

We define an infinite $(0,1)$-matrix $\cH(\Box, \cD)$ as follows:
\begin{gather}
\cH(\Box, \cD)_({i,j}) =
\begin{cases}
1 & \fA_i \Box \fA_j \in \cD \\
0 & \text{otherwise}
\end{cases}
\end{gather}
We denote by $r(\Box, \cD)$ the rank of $\cH(\Box, \cD)_({i,j}$ over $GF(2)$.

Let us look at four binary operations on graphs $G = (V(G), E(G))$ with $E(G) \subseteq V(G)^2$:
Let $G, H$ be such graphs
\begin{enumerate}[(i)]
\item
The disjoint union of $ G= H_1 \sqcup H_2$ of  $H_1$ and $H_2$;
\item
The join union of $G = H-1 \bowtie H_2$ of  $H_1$ and $H_2$, which is the disjoint union of $V(H_1)$ and $V(H_2)$
with 
$$E(G) = e(H_1) \cup E(H_2) \cup \{ e=(i,j : i \in V(H_1), j \in V(H_2) \}.$$
\item
The tensor product of $G = H_1 \times_t H_2$ of  $H_1$ and $H_2$
with $V(G) = V(H_1) \times V(H_2)$ and 
$E(G) = \{ ((u_1, u_2) (v_1, v_2)): (u_1, v_1) \in V(H_1) , (u_2, v_2) \in V(H_2) \}$.
\item
The cartesian product of $G = H_1 \times_c H_2$ of  $H_1$ and $H_2$;
\begin{gather}
E(G) =  \notag \\
\{ ((u_1, u_2) (v_1, v_2)): 
u_1 = v_1  \wedge (u_2, v_2) \in V(H_2) 
\text{  or  }
u_2 = v_2  \wedge (u_1, v_1) \in V(H_1)\}.
\notag
\end{gather}

\end{enumerate}

We compute $\cH( \Box, \cD)$ for $\Box$ any of the above binary operations.
\begin{examples}
\label{ex:lowrank}
\begin{enumerate}[(i)]
\item
A graph $ G= H_1 \sqcup H_2$ is connected iff $H_1$ or $H_2$ is the empty graph.
Hence $r( \sqcup, \cD) =2$.
\item
A graph $ G= H_1 \bowtie H_2$ is connected iff $H_1$ is connected and $H_2$ is the empty graph (or vice versa) or
both are not empty.
Hence $r( \sqcup, \cD) =3$.
\item
A graph $ G= H_1 \times_t H_2$ is connected iff  
one is empty or both are connected and one is non-bipartite.
Hence $r( \sqcup, \cD) =4$. 
\item
A graph $ G= H_1 \times_c H_2$ is connected iff  
one is empty or both are connected.
Hence $r( \sqcup, \cD) =3$. 
\end{enumerate}
\end{examples}

Let $\cD$ be any class of connected graphs. 
There are continuum many such classes,
but there are  only countably many classes of $\tau$-structures which are $\CMSOL$-definable.

\ifskip
\begin{theorem}
\label{th:lowrank}
Let $\Box \in \{\sqcup, \bowtie, \times_t, \times_c\}$.
There are 
continuum many graph properties $\cP$ of graphs with finite $r( \Box, \cP)$.
\end{theorem}
\begin{proof}
Let $A \subseteq \N$ and let $\cP_{conn}(A)$ be the class of connected graphs of order $n \in A$. For $A=\N$ we write $\cP_{conn}$.
\begin{enumerate}[(i)]
\item
There are continuum many properties $\cP_{conn}(A)$.
\item
If $A$ and $A'$ are both not empty proper subsets of $\N$ and $\Box \in \{\sqcup, \bowtie, \times_t, \times_c\}$
$$r(\Box, \cP_{conn}(A)) = r(\Box, \cP_{conn}(A')) = r(\Box, \cP_{conn})=1.$$
\end{enumerate}
\end{proof}
More examples with other binary operations on graphs can be found in \cite[Section 13]{fischer2011application}.

\else
We immediately get from the examples above that there are 
continuum many classes $\cD$ of graphs with $r( \sqcup, \cD) =2$
and continuum many classes $\cD'$ of graphs with $r( \bowtie, \cD') =3$.
More generally we have:

\begin{theorem}
\label{th:lowrank}
For every binary graph operation $\Box$ and every $r$, there are continuum many graph properties $\cP$ with
$r( \Box, \cP) \leq r$.
\end{theorem}
\begin{proof}
Let $\bar{\cP} = \{\cP_i, i \in [m]\}$ be disjoint sequences of graph properties and let $B(x_1, \ldots, x_m)$ be a boolean formula.
For a graph define an assignment $t_i(G)$  of boolean values by setting $t_i(G) =1$ iff $G \in \cP_i$.
We define the graph property $\cP_B$ by $G \in \cP_B$ iff $B(t_1(G), \ldots, t_m(G)) = 1$.
$\cP_B$ depends on the choice of $\bar{\cP}$.

Now let $\Box$ be a binary operation on graphs.
We define the Hankel matrix $\cH(\Box \bar{\cP}, \cP_B)$ by setting
$$
\cH(\Box, \bar{\cP}, \cP_B)_{i,j} = 1 \text{  iff  }
\Box(G_i, G_j) \in \cP_B
$$

{\bf Claim:}  The number of different rows of $\cH(\Box, \bar{\cP}, \cP_B)$ is finite
and bounded by a function of $m$. Hence, the rank of $\cH(\Box, \bar{\cP}, \cP_B)$ is finite. 

Although $\cP_B$ depends on the choice of the graph properties $\cP_i, i \in [m]$,
the Hankel matrix $\cH(\Box, \bar{\cP},  \cP_B)$ does not.
For
$\bar{\cP} = \{\cP_i, i \in [m]\}$ 
and
$\bar{\cP'} = \{\cP'_i, i \in [m]\}$ 
two disjoint sequences of graph properties,  the Hankel matrices coincide, i.e.,
$\cH(\Box, \bar{\cP},  \cP_B) = \cH(\Box, \bar{\cP'},  \cP_B)$.

There are continuum many sequences of disjoint graph properties $\bar{\cP}$.
We conclude that there are continuum many graph properties $\cP_B$ with Hankel matrices $\cH(\Box, \bar{\cP},  \cP_B)$ of rank at most $r(m)$. 
\end{proof}
\fi 

Hankel matrices do occur naturally in Theorem \ref{th:CM-abstract}.

Recall from the introduction the definition of smoothness.
\begin{definition}
\label{def:smooth}
An $n$-ary operation $\Box$ on $\tau$-structures is $\MSOL$-smooth ($\CMSOL$-smooth) if for all $\fA_1, \ldots \fA_n$, $\fB_1, \ldots \fB_n$
such that for all $i \in [n]$ the structures $\fA_i$ and $\fB_i$ satisfy  pairwise the same $\MSOL$ sentences of quantifier rank $q$
then $\Box(\fA_1, \ldots \fA_n)$ and $\Box(\fB_1, \ldots \fB_n)$ also satisfie the same $\MSOL$ ($\CMSOL$) sentences of quantifier rank $q$.
\end{definition}
Both the disjoint union and the join are $\MSOL$-smooth ($\CMSOL$-smooth).
The tensor product and the cartesian product ar not $\MSOL$-smooth (nor $\CMSOL$-smooth),
but they are $\FOL$-smooth, i.e., smooth restricted to First Order Logic $\FOL$.

In \cite{godlin2008evaluations,kotek2014connection} the following was shown:

\begin{theorem}[Finite Rank Theorem]
\label{pr:frank}
Let $\Box$ be a $\MSOL$-smooth or $\CMSOL$-smooth binary operations on $\tau$-structures and let $\cP$ be a $\CMSOL$-definable
graph property.
Then the Hankel matrix $\cH(\Box, \cP)$ has finite rank.
\end{theorem}

\section{$\bF$-circuits and the $\bF$-circuit matrix}
\label{se:circuits}

Let $\tau$ be a finite vocabulary, $\bA$ be a finite set of $\tau$-structures, 
and $\bF$ be a finite set of operations on $\tau$-structures.
In our applications $\tau$ will be usually the vocabulary of graphs augmented by finitely 
many constant symbols and/or finitely many
unary predicates.

\begin{definition}
An {\em inductive $(\bA, \bF)$-class of structures $\cC(\bA, \bF)$} is defined inductively:
\begin{enumerate}[(i)]
\item
$\bA \subseteq \cC(\bA, \bF)$.
For $\fA \in \bA$ its {\em parse tree $t(\fA)$} consists of the single leaf $\fA$.
\item
If $F \in \bF$ is of arity $r$ and $\fA_1, \ldots , \fA_r \in \cC(\bA, \bF)$ with
parse trees $t(\fA_i) i\in [r]$, then
$F(\fA_1, \ldots , \fA_r) \in \cC(\bA, \bF)$ with parse tree 
$F(t(\fA_1), \ldots , t(\fA_r))$. 
\end{enumerate}
\end{definition}

Note that there may be several ways of showing that a $\tau$-structure  $\fA$ is in $\cC(\bA, \bF)$
resulting in different parse trees for $\fA$.

In Section \ref{se:applications} we shall see that the following are 
inductive $(\bA, \bF)$-class of structures:
\begin{enumerate}[(i)]
\item
The class of graphs of tree-width at most $k$;
\item
The class of graphs of clique-width at most $k$;
\item
The class of graphs of modular-width at most $k$;
\item
The class of graphs of twin-width at most $k$;
\end{enumerate}

\begin{definition}
We define $\bF$-circuits inductively:
\begin{enumerate}[(i)]
\item
$x$ is a free leaf.
\item
Every finite $\tau$-structures $\fA$
is a leaf.
\item
Every leaf is a $\bF$-circuit.
\item
If $F \in \bF$ is of arity $r$ and $C_1, \ldots , C_r$ are $\bF$-circuits,
so is $F(C_1, \ldots , C_r)$.
\item
Let $C$ be an $\bF$-circuit and $\fA$ be a $\tau$-structure.
We denote by $C[\fA]$ the $\bF$-circuit obtained from $C$ be replacing all free leaves by the structure $\fA$.
\end{enumerate}
\end{definition}

\begin{definition}
Let $\{C_i: i \in \N^+\}$ be an enumeration of all $\bF$-circuits, and 
$\{\fA_i: i \in \N^+\}$ be an enumeration of all $\tau$-structures.
Furthermore, let $P$ be a class of $\tau$-structures.
\begin{enumerate}[(i)]
\item
The infinite $\bF$-circuit matrix $CM_{\bF,P}$ is defined by
$$
CM_{\bF,P}(i,j) = 
\begin{cases}
1 & C_j[\fA_i] \in P \\
0 & C_j[\fA_i] \not\in P
\end{cases}
$$
\item
$r(CM_{\bF,P})$ denotes the rank of $CM_{\bF,P}$ over $GF(2)$.
\end{enumerate}
\end{definition}

{\bf Proposition \ref{pr:hankel-2}}. 
If $\Box \in \bF$ is a binary operation on $\tau$-structures,  $\cH(\Box, \cP)$ is a submatrix of $\cH(\bA, \bF, \cP)$. 
Furthermore, if $\cH(\bA, \bF, \cP)$ has rank $r$, then $\cH(\Box, \cP)$ has rank $\leq r$.
\begin{proof}
We obtain $\cH(\Box, \cP)$ from $\cH(\bA, \bF, \cP)$ by taking all the columns of the form $\Box(\fA,x)$.
The rank of a submatrix $M'$ of $M$ is always smaller or equal to the rank of $M$. 
Therefore, if $\cH(\bA, \bF, \cP)$ has rank $r$, then $\cH(\Box, \cP)$ has rank $\leq r$.
\end{proof}

{\bf Theorem \ref{th:main}}
Let $\cC$ be a class of $\tau$-structures defined inductively using $\bA, \bF$.
Assume the combined  circuit matrix
$\cH(\bA, \bF, \cP)$ is of finite rank. 
For $\fA \in \cC$ with parse-tree $t(\fA)$, checking whether $\fA \in \cP$ can be doe in linear time
in the size of $t_{\fA}$ where the constants depend only on the rank of $\cH(\bA, \bF, \cP)$. 

\ifskip
\begin{proof}
Since the $\cH(\bA, \bF, \cP)$ has finite rank, it has finitely many different rows [since we work over GF(2)]. 
We select a representative of each.

We want to check whether a structure $\fA$ with parse-tree $t(\fA)$ is in $\cP$. $\fA$ corresponds the root of $t(\fA)$.
We now go over the parse tree, and for each internal node $v$ we find a representative $w$ such that the row of $v$ is the same as the row of $w$.
To do this, suppose that $v$ results from $F(v_1,...,v_r$, $F \in \bF$. 
By induction hypothesis we have
$w_1,...,w_r$ such that $v_i$ and $w_i$ have the same row.
There is a representative $w$ which has the same row as $F(w_1,...,w_r)$ (this is something we can precompute).
It turns out that $v$ and $w$ have the same row (this follows easily from our definitions).
Finally we find a representative which has the same row as the root, and can read off whether $\fA \in \cP$.
\end{proof}
\else
\begin{proof}
Let $\fA$ be a $\tau$-structure and $t(\fA)$ its parse-tree.
The parse-tree can be viewed is a circuit $C(\fA_1, \ldots, \fA_m, \fA)$ without a free leaf.
Let $C_{\fA}(\fA_1, \ldots, \fA_m, x)$ be the circuit where all occurences of $\fA$ are replaced by $x$.
$C(\fA_1, \ldots, \fA_m, \fA)$ is correponds to the entry with row $\fA$ and column $C_{\fA}(\fA_1, \ldots, \fA_m, x)$ in the circuit natrix $\cH(\bA, \bF, \cP)$.
We note that $\fA \in \cP$ iff  the entry  $C(\fA_1, \ldots, \fA_m, \fA)$ has value $1$ in $\cH(\bA, \bF, \cP)$.
All we have to do is 
\begin{enumerate}[(i)]
\item 
to evaluate the circuit $C(\fA_1, \ldots, \fA_m, \fA)$ 
which can be done in linear time in the size of the parse-tree $t(\fA)$, and
\item 
compute its value in the circuit matrix $\cH(\bA, \bF, \cP)$, for which we use that $\cH(\bA, \bF, \cP)$ has finite rank.
\end{enumerate}
\end{proof}
\fi 

\section{Applications}
\label{se:applications}

\subsection{Tree-width}

We define here $\TW(k)$, the class of graphs of tree-width at most $k$ inductively.
This approach is taken from \cite{makowsky2004algorithmic}.
We do it on colored graphs $\bar{G}= (V(G), E(G), C_1, \ldots, C_k)$ augmented with $k$ unary predicates the interpretion of which are disjoint possibly empty sets.
Here, the colors correspond to ports and fusion of ports gives $k$-connections.
The operation $\mathrm{fuse}_i(G)$ contracts the set of elements $C_i(G)$  of $V(G)$ colored with color $i$ to a single point $c$.
If an element $c' \in C_i(G)$ and $(u, c') \in E(G)$  then $(u,c) \in E(\mathrm{fuse}_i(G))$, and the same for $(c', u)$.

\begin{enumerate}[(i)]
\item 
All graphs $\bar{G}$ with at most $k+1$ vertices are in $\TW(k)$. 
\item 
$TW_k$ is closed under disjoint union.
\item 
$TW_k$ is closed under renaming of colours.
\item 
$TW_k$ is closed under {\em fusion}, i.e. for every 
coloured graph $G \in TW_k$,
and for every unary predicate symbol representing a vertex colour $C_i$, 
also $\mathrm{fuse}_i(\bar{G}) \in TW_k$.
\end{enumerate}
This definition is not the standard definition given, 
say in \cite{diestel2000graduate},
but is equivalent to it, cf. \cite{courcelle2002fusion}.

\begin{proposition}
Let $\bar{G}$ be a colored graph with $k$ colors. Then $\bar{G} \in \TW(k)$ iff $\bar{G}$ has tree-width at most $k$.
\end{proposition}
\begin{proof}
In contrast to the standard definition, we have no ports but colors.
We proceed by induction. 
Colored graphs $\bar{G}$ with at most $k+1$ vertices have tree-width at most $k$.
Disjoint union preserves tree-width. Recoloring does not affect tree-width.
Finally, fusion does not increase the tree-width.
Hence all colored graphs in $ \TW(k)$ have tree-width at most $k$.

Conversely, assume $G$ has tree-width at most $k$ and let a tree-decomposition of $G$.
It may be useful to use smooth $k$ tree-decompositions. A $k$ tree-decomposition is smooth 
if every bag has size $k+1$ and the intersection of two neighboring bags has size $k$.
We turn it into a tree decomposition of
$\bar{G}$ by coloring it inductively.
\end{proof}

\begin{theorem}
\label{th:TW-abstract}
\begin{enumerate}[(i)] 
\item
Let $\bar{G} \in \TW(k)$ with a parse-tree $t(\bar{G})$ of size $m(\bar{G})$ and $P$ be a $\CMSOL$-definable property of colored graphs.
Then checking $\bar{G} \in P$ is in linear time  in the size of the parse-stree $m(\bar{G})$.
\item
Let $\bF$ consist of disjoint union, renaming colors $\rho_{i,j}$ and fusion.
Let $P$ be a graph property such that the infinite $\bF$-circuit matrix $CM_{\bF,P}$ is of finite rank, 
then checking $G \in P$ is in linear time  in the size of the parse-stree $m(G)$.
\end{enumerate}
\end{theorem}

\subsection{Clique-width}
We define here $\CW(k)$, the class of graphs of clique-width at most $k$ inductively.
We again define it for colored graphs with at most $k$ colors.
 
\begin{enumerate}[(i)]
\item 
All colored graphs $\bar{G}$ with one vertex colored with oaen of the colors are in $CW_k$.
\item 
$CW_k$ is closed under disjoint union.
\item 
$CW_k$ is closed under renaming of colours.
\item 
$CW_k$ is closed under $\eta_{i,j}(\bar{G})$ which adds all the edges between the two colors $i$ and $j$.
\end{enumerate}

\begin{theorem}
\label{th:CW-abstract}
\begin{enumerate}[(i)] 
\item
Let $\bar{G} \in \CW(k)$ with a parse-tree $t(\bar{G}$ of size $m(\bar{G})$ and $P$ be a $\CMSOL$-definable property of colored graphs.
Then checking $\bar{G} \in P$ is in linear time  in the size of the parse-stree $m(\bar{G})$.
\item
Let $\bF$ consist of disjoint union, renaming colors $\rho_{i,j}$ and 
under $\eta_{i,j}(\bar{G})$ which adds all the edges between the two colors $i$ and $j$.
Let $P$ be a graph property such that the infinite $\bF$-circuit matrix $CM_{\bF,P}$ is of finite rank, 
then checking $G \in P$ is in linear time  in the size of the parse-stree $m(G)$.
\end{enumerate}
\end{theorem}

\subsection{Modular-width}

We consider various substitution operations for $k$-graphs $G = (V(G), E(G), v_1, \ldots , v_k)$.

\begin{definition}
Let $H$ be a graph on $k$ vertices and $G_1, \ldots, G_k$ be graphs $G_i =(V(H_i), E(H_i))$.
The graph $\bar{H} = H[G_1, \ldots, G_k]$ is defined as follows:

\begin{enumerate}[(i)]
\item
$V(\bar{H}) = \bigcup_i^k V(H_i)$.
\item
$E(\bar{H}) = \bigcup_i^k E(H_i) \cup \{(u, v): u \in V(H_i), v \in V(H_j), i \neq j, (v_i, v_j) \in E(G) \}$
\\
Hence $\bar{H} = H[G_1, \ldots, G_k]$ is obtained from $G$
by substituting every vertex $v_i \in V (G)$ with the graph $H_i$ and adding all edges between the
vertices of the graph $H_i$ and the vertices of a graph $H_j$ whenever $(v_i , v_j) \in E(G)$.
\end{enumerate}
We do not require that the vertex sets $V(H_i)$ are disjoint.
\end{definition}

In \cite{gajarsky2013parameterized} an inductive definition of the class of graphs $\MW(k)$ of modular-width at most $k$  is given as follows:
\begin{enumerate}[(i)] 
\item
Graphs consisting of one vertex  are in $\MW(k)$.
\item
If $G_1, G_2 \in \MW(k)$  then $G_1 \sqcup G_2 \in  \MW(k)$, i.e.,  $\MW(k)$
is closed under disjoint unions.
\item
If $G_1, G_2 \in \MW(k)$  then $G_1 \bowtie G_2 \in  \MW(k)$, i.e.,  $\MW(k)$
is closed under the join.
\item
Let $H$ be a graph with exactly $k$ vertices, and let $G_1, \ldots, G_k \in \MW(k)$.
Then $H[G_1, \ldots, G_k] \in \MW(k)$.
\end{enumerate}

\begin{proposition}
\begin{enumerate}[(i)] 
\item
There are only finitely many graphs on $k$ vertices.
Hence there are only finitely many $k$-ary operations
$H[G_1, \ldots, G_k]$.
\item
$H[G_1, \ldots, G_k]$ is $\MSOL$-smooth and also $\CMSOL$-smooth.
\end{enumerate}
\end{proposition}

\begin{theorem}
\label{th:MW-abstract}
\begin{enumerate}[(i)] 
\item
Let $G \in \MW(k)$ with a parse-tree $t(G)$ of size $m(G)$ and $P$ be a $\CMSOL$-definable graph property.
Then checking $G \in P$ is in linear time  in the size of the parse-stree $m(G)$.
\item
Let $\bF$ consist of disjoint union, join, and $H$-substitution for all graphs $H$ on $k$ veatrices.
Let $P$ be a graph property such that the infinite $\bF$-circuit matrix $CM_{\bF,P}$ is of finite rank, 
then checking $G \in P$ is in linear time  in the size of the parse-stree $m(G)$.
\end{enumerate}
\end{theorem}


\section{Conclusions}
\label{se:conclu}

B. Courcelle's famous theorem (Theorem \ref{th:courcelle}) shows that many graph problems which are $\NP$-hard in general,
are in $\FPT$ when restricted to graph classes of bounded tree-width.
This theorem uses logic, more precisely definability in $\MSOL$, as its main hypothesis.
It is therefore a meta-theorem in the sense of \cite{makowsky2025matrix}.
L. Lov\'asz showed in \cite[Theorem 6.48]{lovasz2012large} that the hypothesis of definability in $\MSOL$ can be replaced by a purely combinatorial
hypothesis using connection matrices of finite rank.
A precursor of such a separation may be found in \cite{blatter1984recurrence} where definability in $\MSOL$ can be replaced by substition matrices 
of finite rank.
a special case of Hankel matrices.

In \cite{makowsky2004algorithmic} the second author presented an abstract version of Courcelle's Theorem which still involved logic, stated here as
Theorem \ref{th:CM-abstract}.
Inspired by this we showed 
in this paper how to replace logic in Theorem \ref{th:CM-abstract}.

There are several advantages in replacing definability in logic by a combinatorial condition.
\begin{itemize}
\item
Conceptual: It clarifies the role of logic and of combinatorics. Logic is only used to show that the rank of the circuit matrix for $\cP$ is finite.
\item
Algorithmic: The circuit rank of  $\cP$ is very often very much smaller than what one obtains from definability assumption.
\item
Range of applicability: There are continuum many properties with fixed Hankel rank, whereas there are only countable many $\MSOL$-definable properties $\cP$.
\end{itemize}

However, we do not know how to compute (or even give an upper bound for)  the Hankel rank or the circuit rank for $\cP$ in general.
It remains a challenging project to investigate what one has to know about $\cP$ in order to estimate or compute the Hankel or circuit rank.
Simple examples are given in Examples \ref{ex:lowrank} and Theorem \ref{th:lowrank}.

Courcelle's Theorem was also formulated for numeric graph parameters and graph polynomials. Our approach also extends to this settings.
In this case it is helpful to replace graphs ($\tau$-structures) by formal linear combinations of initely many graphs or structures.
Instead of the field $GF(2)$ one uses now the underlying field or ring of the graph parameter and use as rank the rank over the field or ring.

Recently a new notion of graph width, {\em twin-width} has been defined \cite{bonnet2021twin,bonnet2022twin} and found very useful.
It generalizes most of the currently used notions of width in graph theory.
In particular a logic based version of Courcelle's Theorem is also true for twin-width.
It is, however, not clear how to give a recursive definition of graphs of bounded twin-width which fits our present framework.

\ifskip\else
\fi 
\bibliography{NL}
\bibliographystyle{alpha}
\end{document}